**Crystal orientation and thickness dependence of superconductivity on tetragonal $FeSe_{1-x}$ thin films**

M.J. Wang[1]*, J.Y. Luo[2], T.W. Huang[2], H.H. Chang[3], T.K. Chen[2], F.C. Hsu[1], C.T Wu[3], P.M. Wu[4], A.M. Chang[4], and M.K. Wu[1,3]*,

[1]Institute of Astronomy and Astrophysics, Academia Sinica, Taipei, Taiwan

[2]Institute of Physics, Academia Sinica, Nankang, Taipei, Taiwan

[3]Department of Physics, National Tsing Hua University, Hsinchu, Taiwan

[4]Department of Physics, Duke University, Durham, North Carolina, USA

* To whom Correspondence should be addressed: mingjye@asiaa.sinica.edu.tw and mkwu@phys.sinica.edu.tw

## Abstract

Superconductivity was recently found in the simple tetragonal FeSe structure. Recent studies suggest that FeSe is unconventional, with the symmetry of the superconducting pairing state still under debate. To tackle these problems, clean single crystals and thin films are required. Here we report the fabrication of superconducting β-phase $FeSe_{1-x}$ thin films on different substrates using a pulsed laser deposition (PLD) technique. Quite interestingly, the crystal orientation, and thus, superconductivity in these thin films is sensitive to the growth temperature. At 320°C, films grow preferably along c-axis, but the onset of superconductivity depends on film thickness. At 500°C, films grow along (101), with little thickness dependence. These results suggest that the low temperature structural deformation previously found is crucial to the superconductivity of this material.

Superconductivity was recently discovered in tetragonal $FeSe_{1-x}$ with transition temperature $T_C$ ~8K [1]. Partially substituting relatively larger Te ion to Se site was found to enhance the superconducting transition temperature $T_C$ [2, 3]. It was also found that application of high pressure can even increase the $T_C$ to ~27K [4]. It is believed that more detailed studies of this system, which has the simplest Fe planar structure, may provide critical information to better understand the origin of superconductivity in the iron-based superconductors. Several investigations [5, 6] suggested that $FeSe_{1-x}$ exhibits unconventional superconducting properties. More recently, a detailed NMR study pointed out that there exist interesting electronic characteristics, which are similar to those observed in the electron-doped FeAs superconductors [7]. However, much remains to be answered, such as the pairing symmetry, which must be clearly established, in order to unravel the mechanism of superconductivity in tetragonal $FeSe_{1-x}$. High quality single crystals and thin films are needed in order to tackle these important questions. A notable example of the importance of clean thin films was the identification of d-wave pairing symmetry in high $T_C$ cuprates using superconducting films deposited on tri-crystal substrate [8].

FeSe thin films have been grown on semiconductor substrates to investigate the potential for application in spintronic devices. Most of the reported FeSe thin films were prepared by a selenization process [9-12]. These thin films unfortunately exhibited two or more Fe-Se phases. Shen *et. al.* successfully fabricated tetragonal FeSe thin film with c-axis preferred orientation using low-pressure metal organic chemical vapor deposition (LP-MOCVD) on various substrates [13]. They studied the magnetic and transport properties of these FeSe films but did not observe superconductivity.

Here we report the preparation of superconducting tetragonal $FeSe_{1-x}$ film by the pulsed laser deposition (PLD) technique. The films are found to exhibit preferred orientations with respect to the substrate. It is interesting that the preferred orientation depends on the substrate temperature. At relatively low substrate temperature, the films oriented preferably along (00l), whereas at higher growth temperature the films are along the (l0l) direction. The presence of superconductivity is found to strongly depend on the preferred orientation and the film thickness. These results suggest that the occurrence of superconductivity in $FeSe_{1-x}$ is closely correlated with a low temperature lattice deformation.

The home-made β-$FeSe_{1-x}$ target, 1"ϕ×8mm(t) in size, was prepared by solid state reaction in a vacuum-furnace. The films were deposited on various substrates in vacuum environment (~$10^{-5}$ Torr) using a KrF ($\lambda$ = 248 nm) excimer laser (Lambda Physik LPX Pro). The power density of focused laser on the target was 5~6 J/cm$^2$ and the target-substrate distance was approximately 50 mm. The substrate temperature during deposition varied from 250 °C to 500 °C. We used a ~0.5Å per pulse deposition rate. The film thickness was controlled by the total number of laser pulses and calibrated by surface profiler, α-stepper. The structures of deposited films were preliminarily characterized by an X-ray diffractometer. Subsequent more detailed structure analysis at low temperature were performed using 4-circle X-ray diffractometer with incident beam (12.4 keV) at BL13A in NSRRC and BL12B2 in Spring8, and high resolution transmission electron microscope (HRTEM). The morphologies of films were observed by scanning electron microscopy (SEM). Magnetic susceptibility measurements were performed in a Quantum Design SQUID magnetometer and the temperature dependence of resistance was characterized in a Quantum Design physical property measurement system (QD-PPMS).

We prepared $FeSe_{1-x}$ thin films on MgO substrate with the substrate temperature set at 320$^o$C (hereafter we designate the films grown at this temperature as LT-FeSe films), 420$^o$C, and 500$^o$C (hereafter we designate the films grown at this temperature as HT-FeSe films) The film thickness is estimated close to 100 nm. Figure 1(a) shows the x-ray diffraction patterns for LT-FeSe and HT-FeSe films, respectively. We observed in the LT-FeSe sample only (00$\ell$) peaks of the tetragonal β-phase $FeSe_{1-x}$, suggesting strong c-axis preferred orientation in these films. This result was rather surprising because there is more than 10% lattice mismatch between the β-phase $FeSe_{1-x}$ and the substrates. Figure 1(b) displays the resistance-temperature (R-T) curves of the films deposited at 320°C, 420°C, and 500°C, respectively. It is even more surprising to observe that the LT-FeSe film exhibits only slightly resistive drop at 2K without a complete superconducting transition.

However, it was found to gradually increase the substrate temperature the film properties changed. As shown in Figure 1(a), the X-ray diffraction patterns of HT-FeSe indicate that as the substrate temperature increases the (*00l*) peaks on these films diminish and, instead, the (*l0l*) peaks become dominant. There are also minor peaks belonging to the hexagonal δ-phase on the HT-FeSe films, which is somewhat expected as the FeSe δ-phase is a stable phase at high temperature. Figure 1(b) shows the R-T curves of these films. Clear superconducting transition shows up in the HT-FeSe films.

Detailed transmission electron micrographs for the HT- and LT-FeSe films are shown in Figure 2. The film growth direction is along (*101*) for the HT-FeSe in (a) and (001) for LT-FeSe in (b), revealed by the selected-area electron diffraction (SAED) patterns as shown in the insets. These results are consistent with the X-ray observations. The SAED patterns indicate that both films are well-aligned with MgO substrates. However, a transition layer around 3–6 nm was found in (a) while a much thinner (≤ 1 nm) transition layer is in (b).

The suppression of superconductivity for the LT-FeSe thin film was still puzzling, particularly since the samples showed pure tetragonal phase with nearly epitaxial growth on the substrate. So we next prepared a series of LT- and HT-FeSe with different film thickness. Figure 4a shows the resistive superconducting transition of LT-$FeSe_{1-x}$ films with thickness ranging from 140nm to 1μm. The 140nm film shows a resistive drop around 2K but does not undergo a complete superconducting transition. The onset $T_C$ of film increases with the film thickness, which is shown in the inset of Figure 3a. As the film is made thicker, $T_C$ more closely recovers the bulk result. On the other hand, the HT-FeSe film does not exhibit significant film thickness dependence, as shown in Figure 3b, and inset. The superconducting transition for HT-FeSe is close to the bulk sample value.

The superconductivity in $FeSe_{1-x}$ occurs only in compounds exhibiting Se-deficiency (or Fe-enrichment) [1]. The possibility that the superconductivity disappearing in thin LT-FeSe films, but reappearing in thicker films, is due to the presence of Se-content gradient on the film can be ruled out from the detailed STEM/EDX mapping of the 1 $\mu$m thick LT-FeSe film. The Fe and Se concentrations were found uniformly distributed along the growth direction.

It is generally known that lattice strain may be present at the interface between the thin film and the substrate, especially if the lattice mismatch is large. The strain may subsequently affect the physical properties of the as-grown film. However, the lattice constants derived from the X-ray diffraction of the sample are found very close to the values of the bulk in both LT and HT samples. Thus, it is unlikely for the lattice mismatch to be the dominant cause for the observed strong thickness dependence of superconductivity in the LT-FeSe. This unusual observation led us to investigate in detail whether the low temperature structural distortion observed in the superconducting FeSe could be the cause of such unusual film thickness dependence.

Figure 4(a) is the X-ray diffraction (220) peaks at different temperatures of the superconducting HT-FeSe film (thickness ~150 nm). The results clearly show the splitting of (220) peak confirming the presence of a structural distortion, which occurs around 81.9K. Thus, the lattice strain due to lattice mismatch between the film and the substrate does not affect the structural distortion of HT-FeSe film at low temperature. On the other hand, the LT-FeSe film with 140 nm thickness, which is not superconducting, shows no structural distortion at low temperature as only one single peak appears in the X-ray diffraction pattern at high and low temperatures, Figure 4b top. However, in the diffraction pattern of 1 $\mu$m superconducting film, the (221) diffraction peak gradually becomes broadened as the temperature decreases to 10K, as shown in Figure 4b bottom. This indicates that in LT-FeSe thin film the low temperature distortion is suppressed by the lattice strain, but the strain effect can be gradually released as film thickness increases.

These results can be understood with the following picture. The LT-FeSe film, which exhibits (00l) preferred orientation, would experience stronger confinement effect as the low temperature lattice distortion occurs within the a-b plane. Consequently, the structural distortion in thinner films is suppressed by the lattice strain and results in the disappearance of superconductivity. This strain effect is gradually relaxed as the film thickness increases. On the contrary, the lattice distortion in the HT-FeSe film is not affected by the lattice strain because the film aligns preferably along (l0l) orientation. These results strongly support the conjecture that the dependence of superconductivity on film thickness is associated with the low temperature structural distortion, and further confirm the importance of the low temperature structural distortion to the origin of superconductivity in the FeSe system.

In summary, we have successfully deposited the tetragonal $FeSe_{1-x}$ films on Si (001), MgO, and $SrTiO_3$ substrates by PLD technique. The FeSe film deposited at low temperature, 320 $^{o}$C, reveals highly c-axis orientated characteristic with smooth morphology, but the films prefer to orient along (l0l) instead if prepared at higher growth temperature. Strong thickness dependence of superconductivity is observed in the films with (00l) preferred orientation, but less so in the case with (l0l) preferred orientation. More importantly, we have shown the strong correlation of the low temperature structural distortion to the occurrence of superconductivity in these thin films.

**Acknowledgements**

The authors acknowledge the support from the Taiwan National Synchrotron Radiation Research Center for the detailed low temperature X-ray diffraction. This work was partially supported by the Taiwan National Science Council and the US AFOSR grants.

Figure Captions

**Fig. 1.** (a) The X-ray diffraction pattern of $FeSe_{1-x}$ films deposited at 320°C and 500°C. At higher growth temperature, the intensity of (00l) peaks weakens while the (101) peak intensity becomes stronger. Trace of hexagonal δ-phase FeSe is also present at high temperature. (b) The R-T curves of HT-$FeSe_{1-x}$ films grown at different substrate temperature. At higher growth temperature, the R-T behavior is more metallic-like and a clear superconducting transition is observed. The inset is a closeup of the R-T below 15K.

**Fig. 2.** (a) Cross-section TEM images of 1 $\mu$m (HT-)FeSe thin films grown at 500 °C. The film growth direction is along (*101*), revealed by the selected-area electron diffraction (SAED) patterns (inset). The SAED patterns indicate that the films are well-aligned with MgO substrates. However, a transition layer around 3–6 nm was found. (b) Cross-section TEM images of (LT-)FeSe thin films grown at 320 °C. The film growth direction is along (*001*), revealed by the selected-area electron diffraction (SAED) patterns (inset). The SAED patterns indicate that the films are well-aligned with MgO substrates.

**Fig. 3.** (a) The ρ-T curves of LT-FeSe films with thickness from 150nm to 1030nm. The inset shows the strong thickness dependence of onset $T_C$. As the film is made thicker, Tc approaches the bulk result. (b) The R-T curves of HT-FeSe films with different thickness. The onset $T_C$ shows very little thickness dependence, shown in the inset.

**Fig. 4.** (a) The X-ray diffraction (220) peaks of HT-FeSe film at different temperatures. The results show the splitting of the Bragg peak indicating a structural distortion occurs at around 82K. The inset shows the change of the γ angle with temperature. (b) The X-ray diffraction pattern of LT-FeSe films for (top) 0.15 $\mu$m and (bottom) 1 $\mu$m film at different temperatures. The data show that for 0.15 $\mu$m film the (221) Bragg peak does not change with temperature; for 1 $\mu$m film the (221) peak becomes broadened at low temperature.

Figure 1

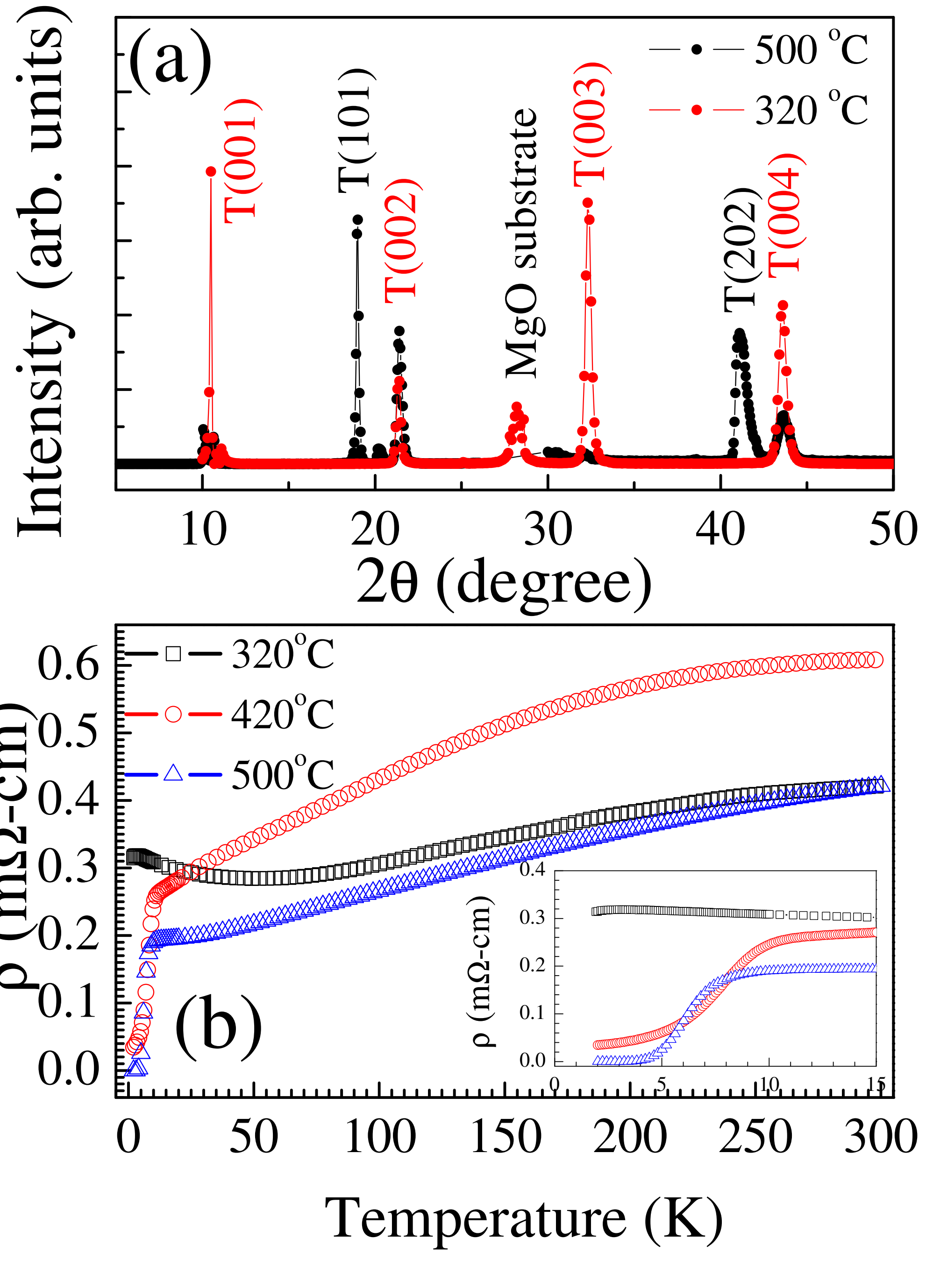

Figure 2

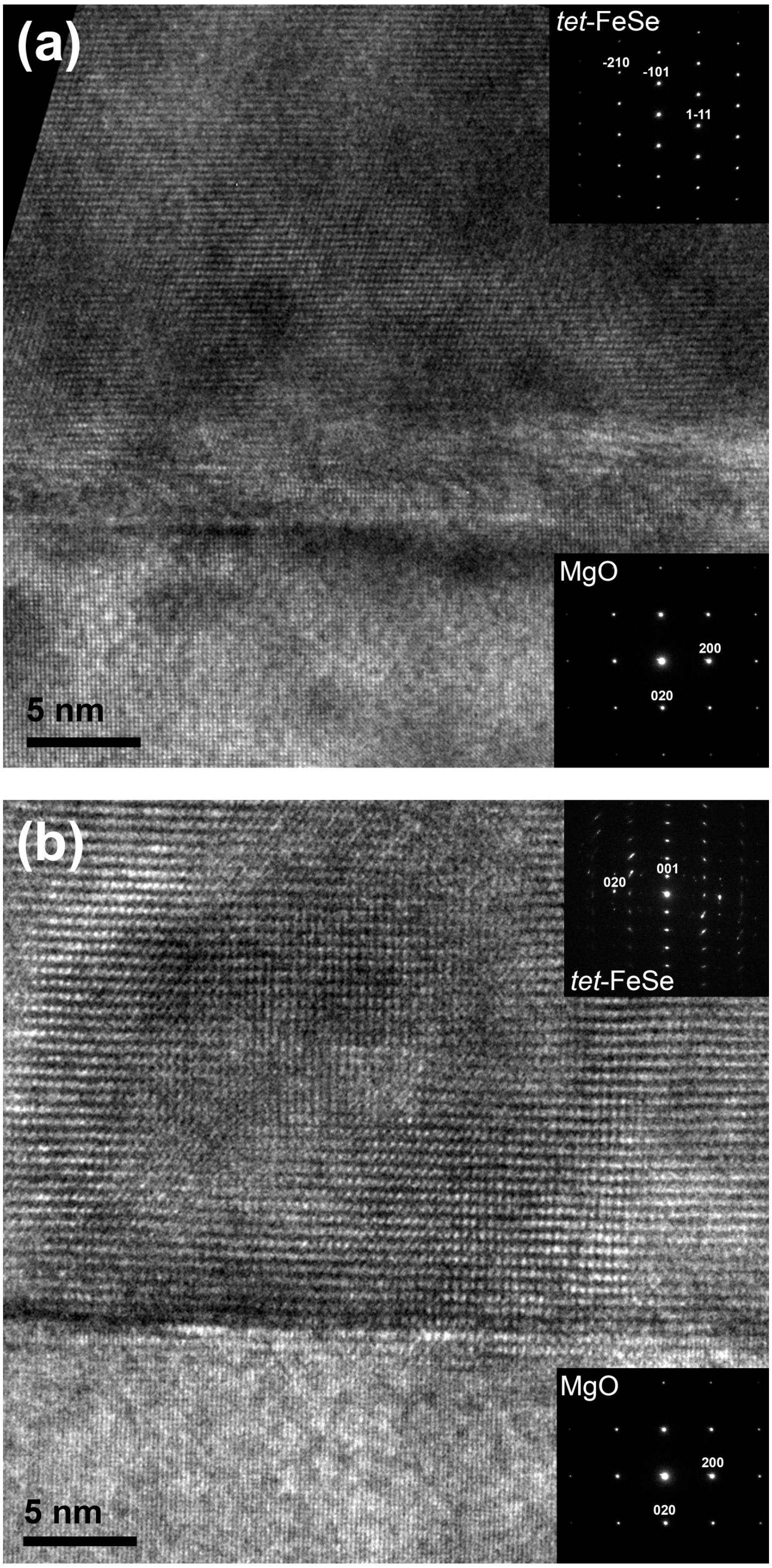

Figure 3

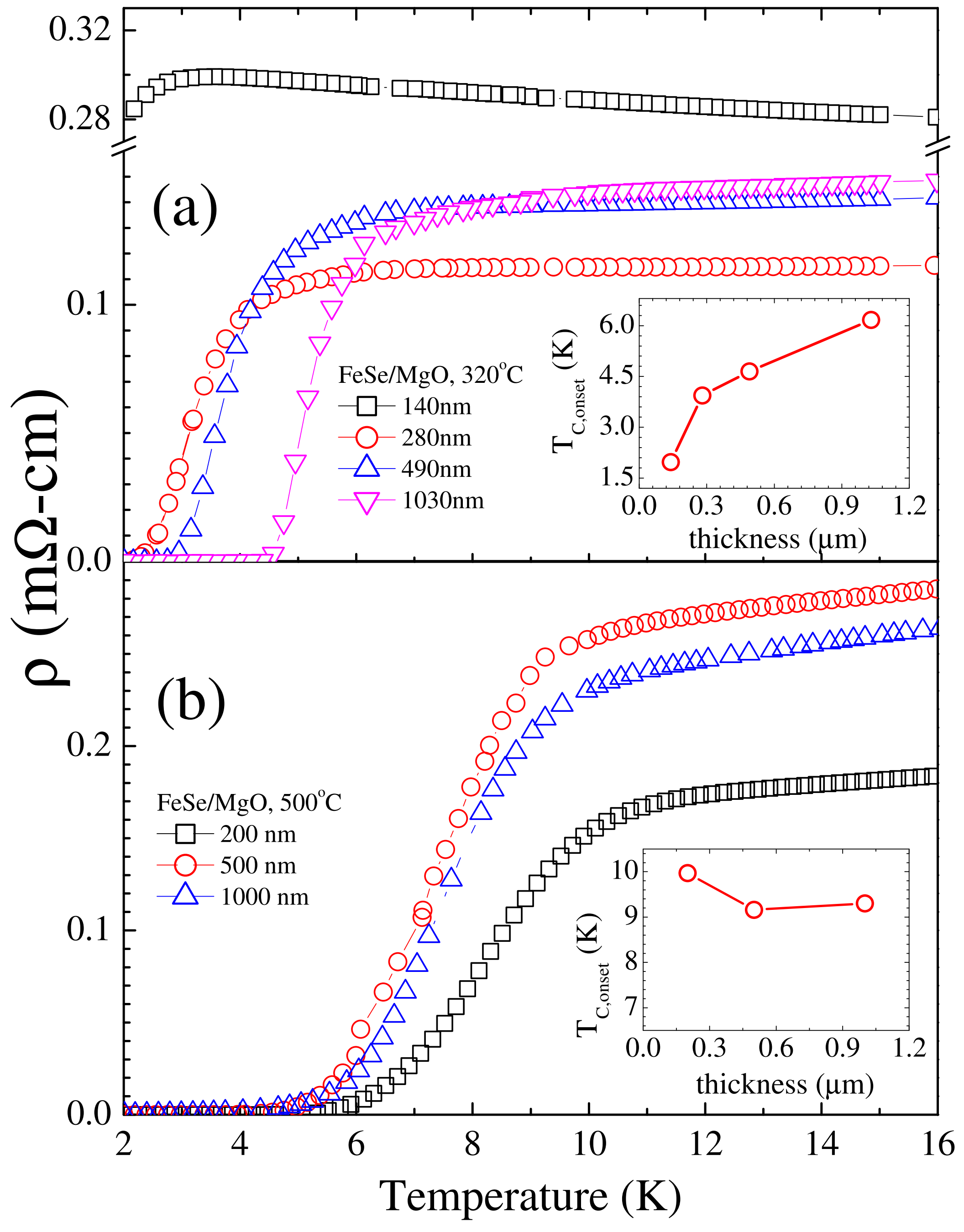

0.32
0.28
(a)
0.1
0.0
FeSe/MgO, 320°C
140nm
280nm
490nm
1030nm
6.0
4.5
3.0
1.5
T C,onset (K)
0.0 0.3 0.6 0.9 1.2
thickness (μm)
ρ (mΩ-cm)
(b)
0.2
0.1
0.0
FeSe/MgO, 500°C
200 nm
500 nm
1000 nm
10
9
8
7
T C,onset (K)
0.0 0.3 0.6 0.9 1.2
thickness (μm)
2 4 6 8 10 12 14 16
Temperature (K)

Figure 4

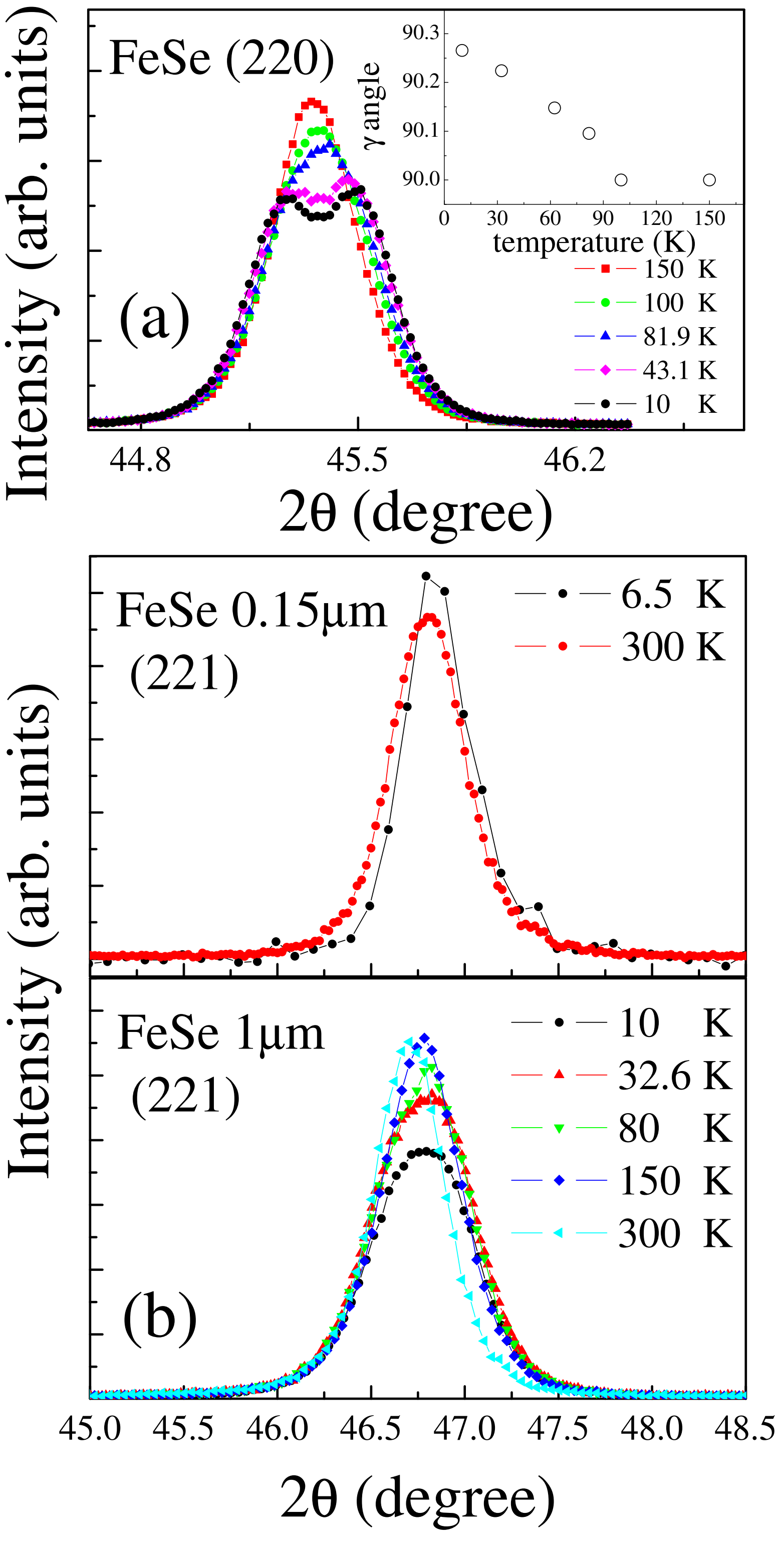